\documentclass[12pt,twocolumn]{iopart}
\usepackage{graphicx}
\usepackage{epsfig}
\usepackage{iopams}
\usepackage{multicol} 
\usepackage{cite}

\newcommand{\kommentar}[1]{}

\newcommand{\ket}[1]{\ensuremath{|#1\rangle}}

\newcommand{\ketbra}[1]{\ensuremath{| #1 \rangle \langle #1 |}}

\newcommand{\eins}{\ensuremath{1\!\!1}}

\newcommand{\BE}{\begin{equation}}
\newcommand{\EE}{\end{equation}}
\newcommand{\be}{\begin{equation}}
\newcommand{\ee}{\end{equation}}
\newcommand{\bea}{\begin{eqnarray}}
\newcommand{\eea}{\end{eqnarray}}
\newcommand{\bean}{\begin{eqnarray*}}
\newcommand{\eean}{\end{eqnarray*}}

\newcommand{\bc}{\begin{center}}
\newcommand{\ec}{\end{center}}

\begin{document}

\title[Recovery of genuine multiparticle entanglement]{Restoring genuine tripartite entanglement under local amplitude damping}

\author{Mazhar Ali\footnote{Email: mazharaliawan@yahoo.com}}
\address{Department of Electrical Engineering, Faculty of Engineering, Islamic University in Madinah, 107 Madinah, Kingdom of Saudi Arabia} 

\begin{abstract}
We investigate the possibility to restore genuine tripartite entanglement under local amplitude damping. 
We show that it is possible to protect genuine entanglement using CNOT and Hadamard gates. We analyze 
several ordering of such recovery operations. We find that for recovery operations applied after exposing qubits to decoherence,
there is no enhancement in lifetime of genuine entanglement. Actual retrieval of entanglement is only possible when reversal scheme 
is applied before and after the decoherence process. We find that retrieval of entanglement for mixture of $|\widetilde{W}\rangle$ state with 
white noise is more evident than the respective mixture of $|W\rangle$ state. We also find the retrieval of entanglement for similar mixture of 
$|GHZ\rangle$ state as well.
\end{abstract}


\section{Introduction}

Entanglement among more than two particles is one of the peculiar features in quantum physics and 
its characterization is an active area of studies \cite{Horodecki-RMP-2009,gtreview}. Pure state entanglement 
is useful for certain tasks like teleportation or cryptography, but in reality unavoidable interactions 
with environment leads to degradation and even abrupt elimination of entanglement from quantum states \cite{Yu-work}. Therefore it is
important to study the techniques to protect entanglement from decoherence. One specific type of decoherence is amplitude 
damping, which is mainly present in ion trap experiments, like atomic qubits subjected to spontaneous emission. Recently, three 
main strategies were proposed to combat amplitude damping. First technique is called weak measurement reversal \cite{Sun-PRA82-2010}, in 
which environment is monitored to restore entanglement probabilistically. Second technique is called quantum measurement reversal 
\cite{Korotkov-PRA81-2010}, where a partial measurement maps a qubit towards ground state before amplitude damping and later another 
measurement restores the initial state. Third scheme is to use Hadamard and CNOT gates to restore a single qubit pure state \cite{Alamri-JPB44-2011}, 
an arbitrary two qubit pure state \cite{Liao-JPB46-2013}, and mixed states of two qubits in a weak measurement \cite{Esfahani-arXiv-2016}.

Effects of decoherence on entanglement among multiparticle states have extensively been studied and this is an active area of research 
\cite{lifetime, acindec, bipartitedec,lowerbounds, Guehne-PRA78-2008, furtherdec, Carnio-PRL115-2015, Ali-JPB47-2014}. 
Several works considered the life time of entanglement under decoherence \cite{lifetime, acindec}. 
Here, the life time of entanglement denotes the time with a nonzero value of the chosen measure of entanglement. Some works studied bipartite 
aspects of the entanglement of several particles \cite{bipartitedec}, however it gives a partial description, because multiparticle entanglement 
is different from entanglement between all bipartitions \cite{gtreview}. Another main problem behind previous studies is lack of availability of 
a fully developed theory of multiparticle entanglement. Hence, one could only make statements about lower bounds on entanglement instead of its 
actual value \cite{lowerbounds}. The exact calculation of a multiparticle entanglement measure was possible for special states and decoherence 
models \cite{Guehne-PRA78-2008}. Recent progress in the theory of multiparticle entanglement, especially the computable entanglement monotone 
for genuine multiparticle entanglement from Ref. \cite{Bastian-PRL106-2011}, enabled us to study the effects of decoherence on multiparticle 
entanglement \cite{Ali-JPB47-2014}. 

In our current study, we utilize the computable genuine negativity \cite{Hofmann-JPA47-2014} to analyze the possibility to restore 
genuine entanglement between three qubits undergoing local amplitude damping. We investigate the effects of applying local recovery operations, 
which are composed of Hadamard and CNOT gates, in several ordering to find the optimum results. We find that optimum restorage of genuine 
entanglement is only possible for the case, when the recovery operations are applied both before and after the decoherence process. 
In all other cases, the 
life time and amount of genuine entanglement is not restored considerably. We analyze the mixtures of W-type states and GHZ-type states with 
white noise (maximally mixed state). We find that locally equivalent state to W-state is more restored in comparison with W-state, because the 
error matrix has zero entries for off-diagonal elements for $|\widetilde{W}\rangle$ state. Although we talk about multipartite quantum systems and 
multiparticle entanglement, however we study three two-level quantum systems (or three qubits). As the discussion on genuine entanglement is general 
and applicable for an arbitrary finite dimensional quantum system (quNit) and finite number of parties (M), this means that results are applicable to 
multi-qubit systems. 

This work is organized as follows. In Section \ref{Sec:Model}, we describe amplitude damping model to obtain the dynamics 
of an arbitrary density matrix. In Section \ref{Sec:GME}, we briefly review the concept of multiparticle entanglement and also review the 
derivation of genuine negativity. In Section \ref{Sec:Rec}, we define the investigated recovery schemes. 
We present the main results in Section \ref{Sec:Results}. Finally, we offer some conclusions in Section \ref{Conc}.

\section{Preliminaries}

\subsection{Local amplitude damping model for multiparticle states} 
\label{Sec:Model}

We consider $N$ qubits (e.g., $N$ two level atoms) which are coupled to their own local reservoirs. The reservoirs are 
assumed to be independent from each other. We assume weak coupling between each qubit and the corresponding reservoir and no back action 
effect of the qubits on the reservoirs. We also assume that the correlation time between the qubits and the reservoirs is much shorter 
than the characteristic time of the evolution so that the Markovian approximation is valid. The interactions of the physical system with 
environment is usually studied either by solving a master equation, using the Kraus operator formalism, and quantum trajectories. 
We work in the Kraus operator formalism. The time evolution of an initial density matrix can be written as
\begin{equation}
\varrho(t) = \sum_{i=1}^{2^N}  \, K_i(t) \, \varrho(0) \, K_i^\dagger(t), 
\label{Eq:TE}
\end{equation}
where $K_i(t)$ are the Kraus operators, satisfying the normalization condition $\sum_i \, K_i^\dagger(t) \, K_i(t) =  \eins$ and $N$ is the number 
of qubits. The precise form of these Kraus operators are given as 
$ K_i (t) = \omega_{i_1}^A  \otimes \omega_{i_2}^B \otimes \cdots \otimes \omega_{i_N}^N, $
where $\omega_i^j$ are the single-qubit Kraus operators acting on the $j$th qubit. 

For {\it amplitude damping}, there are two Kraus operators for a single qubit,
\be
\omega_1^j = \left( 
\begin{array}{cc}
1 & 0 \\ 
0 & \gamma_j
\end{array}
\right) \, , \quad \omega_2^j = \left( 
\begin{array}{cc}
0 & \sqrt{1-\gamma_j^2} \\ 
0 & 0
\end{array}
\right), 
\ee
where $\gamma_j = {\rm e}^{- \Gamma_j t/2}$ and $\Gamma_j$ is the spontaneous emission decay rate of $j$th qubit. In ion-trap experiments, 
amplitude damping is the typical noise. For the case of $3$-qubit states, there are $2^3$  global Kraus operators $K_i(t)$ for the amplitude 
damping channels. For the sake of simplicity we assume onwards that $\gamma_A = \gamma_B = \cdots = \gamma_N = \gamma$.

The time evolved density matrix for a single qubit can directly be computed. Under amplitude damping it is given as 
\be
\varrho (t) = \left( \begin{array}{cc}
\varrho_{11} + \varrho_{22} (1-e^{- \gamma t}) & \varrho_{12} \, e^{- \gamma t/2 } \\ 
\varrho_{21} \, e^{- \gamma t/2 } & \varrho_{22} \, e^{- \gamma t} 
\end{array} \right),
\ee
where $\varrho_{ij}$ are initial density matrix elements. For more qubits, the calculation of density matrices is straightforward.

\subsection{Genuine multiparticle entanglement and genuine negativity} 
\label{Sec:GME}

We review the basic definitions of genuine entanglement and genuine negativity by taking three 
parties $A$, $B$, and $C$, with generalization to more parties as straightforward. A state is separable with respect 
to some bipartition, say, $A|BC$, if it is a mixture of product states with respect to this partition, that is, 
$\rho = \sum_j \, p_j \, |\psi_A^j \rangle\langle \psi_A^j| \otimes |\psi_{BC}^j \rangle\langle \psi_{BC}^j|$, 
with $p_j$ a probability distribution. We name these states as $\rho_{A|BC}^{sep}$. Similarly, we can write other states as 
$\rho_{B|CA}^{sep}$ and $\rho_{C|AB}^{sep}$. A state is biseparable if it is convex combination of these states, that is 
\begin{eqnarray}
 \rho^{bs} = q_1 \, \rho_{A|BC}^{sep} + q_2 \, \rho_{B|CA}^{sep} + q_3 \, \rho_{C|AB}^{sep}\,,
\end{eqnarray}
with $\sum_i q_i = 1$. Finally, a state is genuinely entangled if not biseparable. In this paper, we always mean genuine 
multipartite entanglement when we talk about entanglement. 

Meantime a technique has been developed to detect and characterize multipartite entanglement \cite{Bastian-PRL106-2011}. The method is based on 
using positive partial transpose (PPT) mixtures. A bipartite state 
$\rho = \sum_{ijkl} \, \rho_{ij,kl} \, |i\rangle\langle j| \otimes |k\rangle\langle l|$ is PPT if its partially transposed matrix 
$\rho^{T_A} = \sum_{ijkl} \, \rho_{ji,kl} \, |i\rangle\langle j| \otimes |k\rangle\langle l|$ has no negative eigenvalues. 
The separable states are always PPT \cite{peresppt}. The set of separable states with respect to some partition 
is therefore contained in a larger set of states which has a positive partial transpose for that bipartition. 
The PPT states with respect to some bipartition are $\rho_{A|BC}^{PPT}$, $\rho_{B|CA}^{PPT}$, and $\rho_{C|AB}^{PPT}$. 
If a state can be written as
\begin{eqnarray}
\rho^{PPTmix} = r_1 \, \rho_{A|BC}^{PPT} + r_2 \, \rho_{B|CA}^{PPT} + r_3 \, \rho_{C|AB}^{PPT}\,, 
\end{eqnarray}
it is called a PPT mixture. Here $\sum_i r_i = 1$. Any biseparable state is a PPT mixture, hence any state which is not a PPT mixture is guaranteed 
to be genuinely entangled. The major advantage of taking PPT mixtures instead of biseparable states is the fact that PPT mixtures can be fully 
characterized by the method of semidefinite programming (SDP), a standard  method in convex optimization \cite{sdp}. Generally the set of PPT mixtures 
is a very good approximation to the set of biseparable states and delivers the best known separability criterion for many cases; however, there are 
genuine entangled states which are PPT mixtures \cite{Bastian-PRL106-2011}.

It was proved \cite{Bastian-PRL106-2011} that a state is a PPT mixture iff the optimization problem 
\begin{eqnarray}
\min {\rm Tr} (\mathcal{W} \rho)
\end{eqnarray}
with constraints that for all bipartition $M|\bar{M}$
\begin{eqnarray}
\mathcal{W} = P_M + Q_M^{T_M},
 \quad \mbox{ with }
 0 \leq P_M\,\leq 1 \mbox{ and }
 0 \leq  Q_M  \leq 1\, 
\end{eqnarray}
has a positive solution. Here $T_M$ means partial transpose with respect to partition $M$. The constraints means that the operator $\mathcal{W}$ is a 
decomposable entanglement witness for any bipartition. For a negative minimum, $\rho$ is not a PPT mixture and hence is genuinely entangled. 
For the semidefinite program (SDP), the minimum can be efficiently computed and the optimality of the solution can be certified \cite{sdp}. 
For solving the SDP we used the programs YALMIP and SDPT3 \cite{yalmip}, an implementation which is freely available \cite{pptmix}.

This approach can be used to quantify genuine entanglement and the absolute value of this minimization was proved to be an entanglement monotone 
for genuine entanglement \cite{Bastian-PRL106-2011, Hofmann-JPA47-2014}. We denote this measure as genuine negativity $E(\rho)$ or 
$E$-monotone \cite{Hofmann-JPA47-2014}. 
For bipartite systems, this monotone is equivalent to {\it negativity} \cite{Vidal-PRA65-2002}. For a system of qubits, this measure is bounded by 
$E(\rho) \leq 1/2$ \cite{Bastian-PRL106-2011}.

\subsection{Recovery schemes}
\label{Sec:Rec}

The basic recovery scheme, which we utilize to study our problem consist of Hadamard and CNOT gates. It was shown that an arbitrary single 
qubit state can be completely recovered using this technique \cite{Alamri-JPB44-2011}. Later on, this method was generalized to restore two 
qubits quantum states and entanglement \cite{Liao-JPB46-2013, Esfahani-arXiv-2016}. 
In our work, we analyze several scenarios where these local operations are performed on an arbitrary quantum state of three qubits. We extend the 
previous studies in two ways. First, we study this scheme for multipartite quantum systems and second we study the various ordering of applying 
these recovery operations and their effect on restorage of genuine entanglement. 

\begin{figure}[t!]
\scalebox{2.25}{\includegraphics[width=2.0in]{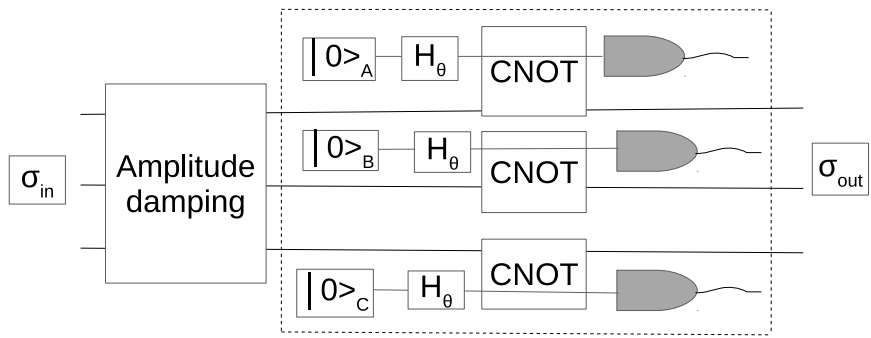}}
\centering
\caption{Circuit diagram for recovery operations to restore genuine entanglement for three-qubit states under amplitude damping. 
See text for further details.}
\label{Fig:Model1}
\end{figure}

The recovery schemes can be outlined as follows. First, we take an arbitrary initial quantum state $\sigma_{in}$. In addition to initial state, we 
take three auxiliary qubits which are in $|000\rangle$ state. To create superpositions of ground and excited state, a Hadamard gate is applied 
on each auxiliary qubit individually, such that 
\be
\rho_{A_i} = H_\theta ( |0\rangle\langle0| ) H_\theta^\dagger =  \left( \begin{array}{cc}
\cos^2 \theta & \cos \theta \sin \theta \\ 
\cos \theta \sin \theta & \sin^2 \theta 
\end{array} \right),
\ee
where 
\be
H_\theta =  \left( \begin{array}{cc}
\cos \theta & - \sin \theta \\ 
\sin \theta & \cos \theta 
\end{array} \right).
\ee
After this step, three CNOT gates are separately applied to each pair of system qubit in state $\sigma_{in}$ and auxiliary qubit with system qubit as 
control qubit and auxiliary qubit as target qubit. After this procedure, we take measurements on auxiliary qubits. If the measurement results are 
$|000\rangle$, then the recovery process is successful otherwise fails. The probability of this success process can be worked out easily, which depends 
on when we choose to apply these operations. The dashed box in Figure~(\ref{Fig:Model1}) depicts the essential recovery operations. The final state after 
taking measurements on auxiliary qubits is denoted by $\sigma_{out}$.

First scheme, we focus on the case when recovery operations are applied only for one time and that after the initial state $\sigma_{in}$ is exposed to local 
amplitude damping. This situation is shown in the circuit diagram of Figure~(\ref{Fig:Model1}). We will show the effects of 
this scheme on genuine entanglement in next section. Here we just want to mention that similar to case of two qubits 
\cite{Liao-JPB46-2013, Esfahani-arXiv-2016}, here these operations are unable to affect the dynamics of genuine entanglement in a considerable 
way. Only the numerical value of entanglement is slightly enhanced. This means that it is not possible to delay the life time of entanglement 
with these operations and genuine negativity vanishes at the same time as it vanishes without any recovery operations. For this scheme, we choose 
$\tan \,\theta = 1/\gamma$ to compensate for decoherence effects.

\begin{figure}[t!]
\scalebox{2.25}{\includegraphics[width=2.0in]{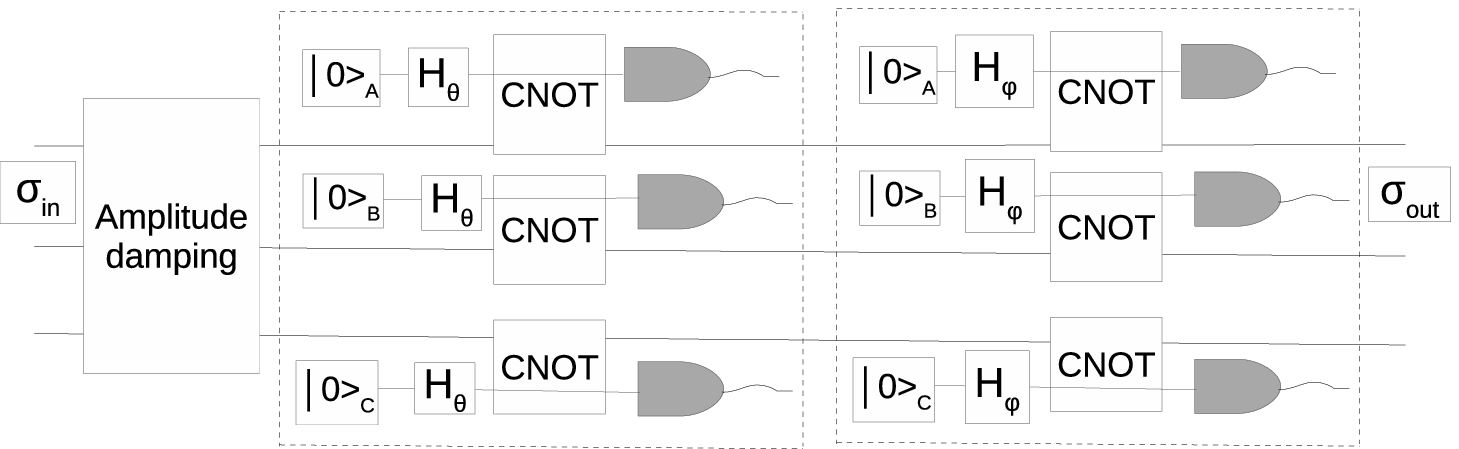}}
\centering
\caption{Circuit diagram for two consequetive operations to restore genuine entanglement for three-qubit states under amplitude damping. See text
for further details.}
\label{Fig:Model3}
\end{figure}
Another question, we can ask is that what happens if we apply two consequetive recovery operations after the qubits are 
exposed to amplitude damping channels. Figure~(\ref{Fig:Model3}) depicts this scheme. We will present the detailed results in the 
next section but an interesting feature is the observation that if we choose the relation 
``$\tan \theta \, \tan \phi \, \gamma = 1$'', then applying recovery operations is equivalent to applying it for single time. 
Here $\theta$ ($\phi$) denotes rotation angle in Hadamard gate for first (second) recovery operations. It means any number of 
such operations satisfying this condition do not affect the dynamics at all. Whereas if we choose $\tan \phi = 1/\gamma$ and keep 
$s = \tan \theta$, then still it is not possible to delay life time of entanglement. We discuss the detailed results in next section. 
 
\begin{figure}[t!]
\scalebox{2.25}{\includegraphics[width=2.0in]{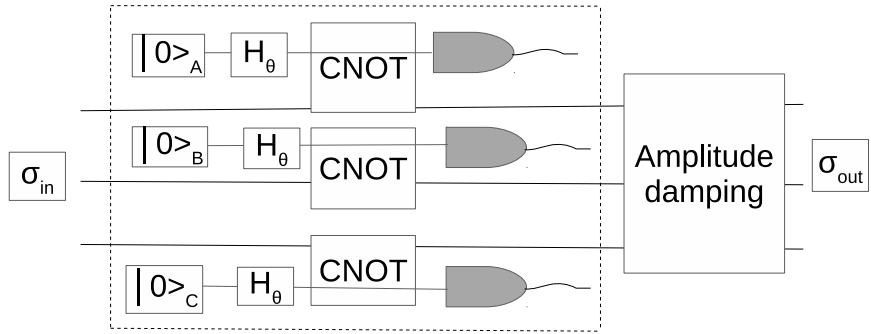}}
\centering
\caption{Circuit diagram for single recovery operations applied before the three-qubit states are exposed to amplitude damping. See text
for further details.}
\label{Fig:Model4}
\end{figure}
We have seen that by the above mentioned order of applying recovery operations, we do not get our aim, that is to enhance the life time of genuine 
entanglement against amplitude damping. Therefore, we apply recovery operations for single time, in order to make the initial state robust 
against amplitude damping. After this process, we expose the qubits to decoherence. The circuit diagram is shown in Figure (\ref{Fig:Model4}). 
We will see in next section that the life time of genuine entanglement is enhanced in this case upto some extent.

\begin{figure}[t!]
\scalebox{2.25}{\includegraphics[width=2.0in]{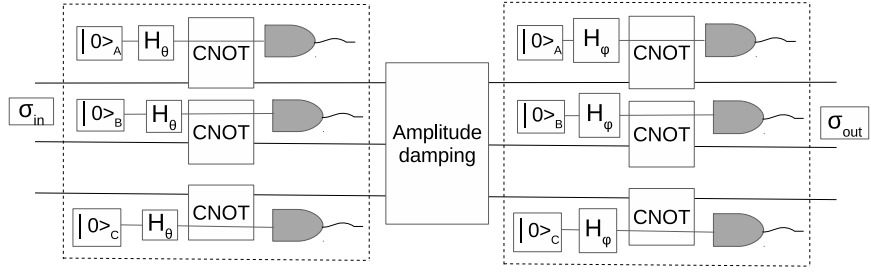}}
\centering
\caption{Circuit diagram for double recovery operations applied before the three-qubit states are exposed to amplitude damping and once afterwards. 
See text for further details.}
\label{Fig:Model2}
\end{figure}

Finally we consider the situation where we apply the recovery operations before and after the qubits undergo amplitude damping. We 
will see in next section that only in this case the entanglement is protected considerably. So this scheme is very effective in recovering 
genuine entanglement against amplitude damping. Figure (\ref{Fig:Model2}) depicts the circuit diagram for this case. In this case, we 
have chosen the parameters such that $\tan \phi = 1/(x \, \gamma)$, where ``$x$'' denotes the parameter for first recovery operations.
 
We have observed that for schemes shown in Figure (\ref{Fig:Model1}) and Figure (\ref{Fig:Model2}), it is always possible to write the final 
density matrix $\rho_f$  as sum of an initial state and an error matrix, that is,
\be
\rho_f = \frac{1}{\mathcal{N}} \, \big[ \, \rho_i + \rho_{err} \, \big]\,,
\label{Eq:REC}
\ee
where $\mathcal{N}$ is the normalization factor and $\rho_{err}$ is the error matrix given as
\be
\rho_{err} = \left( 
\begin{array}{cccccccc}
\tilde{\rho}_{11} & \tilde{\rho}_{12} & \tilde{\rho}_{13} & \tilde{\rho}_{14} & \tilde{\rho}_{15} & \tilde{\rho}_{16} & \tilde{\rho}_{17} & 0 \\ 
\tilde{\rho}_{21} & \tilde{\rho}_{22} & \tilde{\rho}_{23} & \tilde{\rho}_{24} & \tilde{\rho}_{25} & \tilde{\rho}_{26} & 0 & 0 \\
\tilde{\rho}_{31} & \tilde{\rho}_{32} & \tilde{\rho}_{33} & \tilde{\rho}_{34} & \tilde{\rho}_{35} & 0 & \tilde{\rho}_{37} & 0 \\
\tilde{\rho}_{41} & \tilde{\rho}_{42} & \tilde{\rho}_{43} & \tilde{\rho}_{44} & 0 & 0 & 0 & 0 \\
\tilde{\rho}_{51} & \tilde{\rho}_{52} & \tilde{\rho}_{53} & 0 & \tilde{\rho}_{55} & \tilde{\rho}_{56} & \tilde{\rho}_{57} & 0 \\
\tilde{\rho}_{61} & \tilde{\rho}_{62} & 0 & 0 & \tilde{\rho}_{65} & \tilde{\rho}_{66} & 0 & 0 \\
\tilde{\rho}_{71} & 0 & \tilde{\rho}_{73} & 0 & \tilde{\rho}_{75} & 0 & \tilde{\rho}_{77} & 0 \\
0 & 0 & 0 & 0 & 0 & 0 & 0 & 0 \\
\end{array}
\right) \, , 
\label{Eq:ERR}
\ee
where $\tilde{\rho}_{ij}$ are error matrix entries. The error matrix in both situations have exactly same zero matrix elements. The nonzero 
matrix elements depends on the type of recovery scheme and have different expressions for each recovery scheme. 
$\rho_i$ is an arbitrary initial state of three qubits. It is interesting to note that for locally equivalent state to $W$-state given as
\be
|\widetilde{W}\rangle = \frac{1}{\sqrt{3}} \, (|110 \rangle + |101\rangle + |011 \rangle)\,,
\label{Eq:WT}
\ee
has zero entries in error matrix, that is, $\tilde{\rho}_{46} = \tilde{\rho}_{47} = \ldots = \tilde{\rho}_{76} = 0$. This means that for an initial 
$|\widetilde{W}\rangle$ state, the error matrix would have minimum entries, and better recovery of quantum state and entanglement.

In rest of this work, by $\rho_d$, we mean density matrix under decoherence without any type of recovery operations. 
By $\rho_r$, we mean the recovery scheme applied only once and that after exposing qubits to amplitude damping (Figure (\ref{Fig:Model1})). 
By $\rho_{dr}$, we mean double recovery scheme in which recovery operations are applied before and after the qubits are exposed to local amplitude 
damping (Figure (\ref{Fig:Model2})).

\section{Results} 
\label{Sec:Results}

It is well known that for three qubits, there exist two inequivalent genuine multiparticle entangled states, which can not be transformed into 
each other by any SLOCC \cite{gtreview}, namely the $GHZ$ states and the $W$ states, 
\begin{eqnarray}
|GHZ \rangle &=& \frac{1}{\sqrt{2}}(\ket{000} + \ket{111}),
\nonumber
\\
\ket{W}&=&\frac{1}{\sqrt{3}}(\ket{001} + \ket{010} +\ket{100}).
\label{Eq:GHZ3Qb1}
\end{eqnarray}
For the GHZ state, genuine negativity has a value of $E(\ketbra{GHZ}) = 1/2$, while for the W state, its value is
$E(\ketbra{W}) \approx 0.443$. We consider the effects of applying recovery schemes on mixture of GHZ state with maximally mixed state, given as
\be
\rho_{GHZ} = \alpha \, |GHZ\rangle\langle GHZ| + \frac{1-\alpha}{8} \, I_8,
\label{Eq:GHZM}
\ee
where $0 \leq \alpha \leq 1$ and $I_8$ is $8 \times 8$ identity matrix. The entanglement properties of these states are 
well known and entanglement criterion discussed in previous section is also necessary and sufficient for detection of 
genuine entanglement of such mixtures \cite{Bastian-PRL106-2011}. Similarly, we can write the equation for mixture of $W$ state as  
\be
\rho_W = \alpha \, |W\rangle\langle W| + \frac{1-\alpha}{8} \, I_8,
\label{Eq:WM}
\ee
We set $\alpha = 0.8$ for these class of states in this work to compare the behaviour of recovery operations. 

\begin{figure}[t!]
\scalebox{2.25}{\includegraphics[width=1.99in]{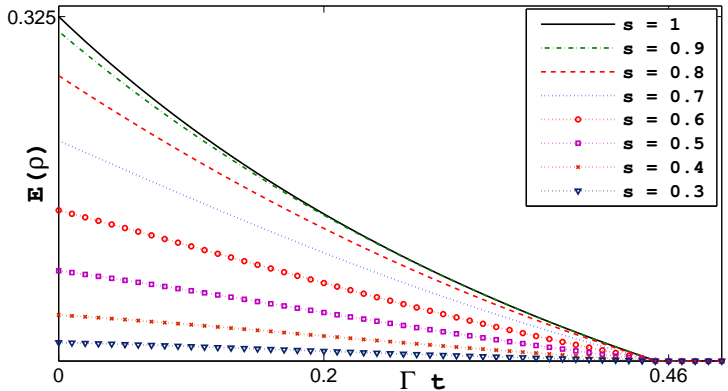}}
\centering
\caption{Entanglement monotone plotted against parameter $\Gamma \, t$ for states Eq.(\ref{Eq:GHZM}) with $\alpha = 0.8$ and 
various values of $s$.}
\label{Fig:E3QGHZRv1}
\end{figure}

As discussed earlier, if qubits are first exposed to amplitude damping and after that we apply recovery operations for a single time then 
there is no effect on the life time of genuine entanglement except that its numerical value is slightly enhanced. 
On the other hand, if we apply the recovery operations two times after decoherence (see Figure~(\ref{Fig:Model3})), then 
the effects on dynamics depends on $\tan \theta$ (rotation angle for first recovery operation) and $\tan \phi$ (rotation angle 
for second recovery operation). For the choice that meet the condition ``$\tan \theta \, \tan \phi \, \gamma = 1$'', we find that dynamics 
is not affected at all, and second operations acts as identity matrix. However, if we set ``$s = \tan \theta$'', and choose 
``$\tan \phi \, \gamma = 1$'', then the dynamics is quite different for each value of ``$s$''.  Figure~(\ref{Fig:E3QGHZRv1}) shows 
genuine negativity plotted against the parameter $\Gamma \, t$ for various values of ``$s$''. We observe that for lower values 
of ``$s$'', the amount of initial entanglement also becomes lower, however, interestingly each curve comes to an end at a same point 
$\Gamma \, t \approx 0.46$, irrespective of initial amount of genuine entanglement. 

\begin{figure}[t!]
\centering
\scalebox{2.25}{\includegraphics[width=1.99in]{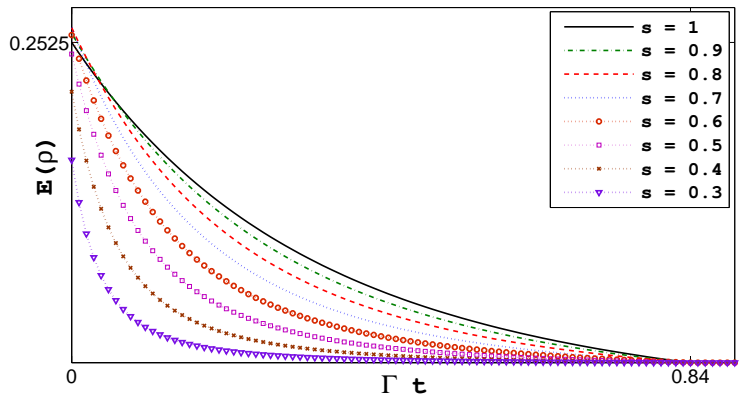}}
\caption{Entanglement monotone plotted against parameter $\Gamma \, t$ for states Eq.(\ref{Eq:WM}) with $\alpha = 0.8$ and 
various values of $s$.}
\label{Fig:E3QWRv1}
\end{figure}

Figure~(\ref{Fig:E3QWRv1}) shows similar results for an initial mixture of $W$ states. Once again we see the same trend as in the case of 
$GHZ$ states except that decay is relatively faster. For all choices of parameter ``$s$'', genuine entanglement ends at $\Gamma \, t \approx 0.84$. 
So we conclude that all such schemes in which recovery operations are performed after the qubits are exposed to amplitude damping can not 
enhance the life time of genuine entanglement.

Let us now focus on the second category of recovery scheme, where we apply recovery operations before qubits are allowed to 
interact with the environment. This scheme can be further divided into two cases. In the first case, we apply recovery operations only 
initially and then let the qubits undergo decoherence process. In second case, we apply recovery operations before and after the 
qubits are exposed to amplitude damping noise (see Figures (\ref{Fig:Model4}) and (\ref{Fig:Model2})). We find out that only in this 
category, the life time of genuine entanglement can be enhanced considerably.

\begin{figure}[t!]
 \scalebox{2.25}{\includegraphics[width=1.99in]{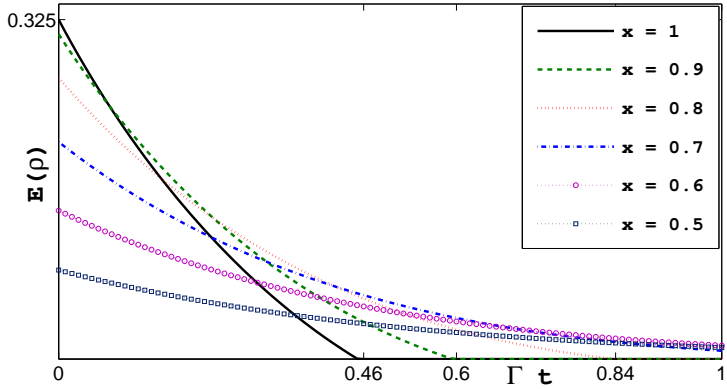}}
 \centering
 \caption{Genuine negativity for mixture of three-qubit $GHZ$ states is plotted against parameter $\Gamma \, t$ and various values of 
 parameter ``$x$''.}
 \label{Fig:R3QbGHZc2}
 \end{figure}

Figure~(\ref{Fig:R3QbGHZc2}) depicts the results for $GHZ$ mixture for various values of parameter ``$x = \tan \theta$''. We conclude 
two things from this figure. First, decreasing the value of ``$x$'' prolongs the life time of genuine entanglement and second the actual initial 
amount of genuine entanglement also decreases by decreasing the parameter ``$x$''. However, we should bear in mind that decreasing parameter 
``$x$'' means the decreasing probability for the successful recovery of state and protocol. However, we see that this scheme can delay the 
vanishing of genuine entanglement with some compromise on failure of the recovery protocol.  

\begin{figure}[t!]
\centering
\scalebox{2.25}{\includegraphics[width=1.99in]{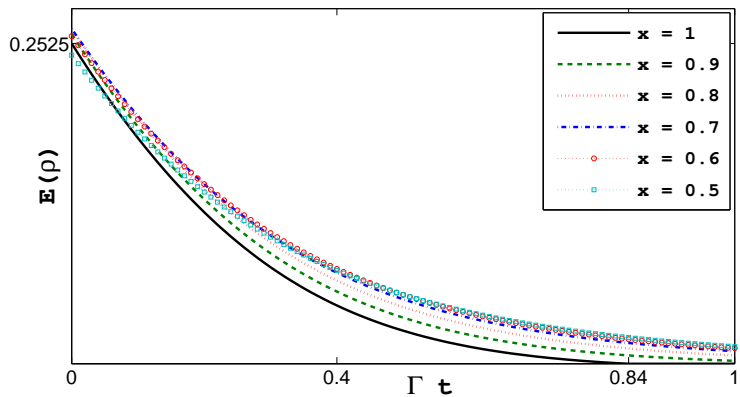}}
\caption{Genuine negativity for mixture of three-qubit $W$ states is plotted against parameter $\Gamma \, t$ and various values of 
 parameter ``$x$''.}
\label{Fig:R3QbWc2}
\end{figure}

Figure~(\ref{Fig:R3QbWc2}) shows the results for similar mixture of $W$ state. We see similar trend in the prolongation of life time of 
genuine entanglement, however as compared with mixtures of $GHZ$ state, the amount of initial entanglement do not vary much. Actually 
for some values of parameter ``$x$'', the initial entanglement increases slightly as well. 

\begin{figure}[t!]
\scalebox{2.25}{\includegraphics[width=1.99in]{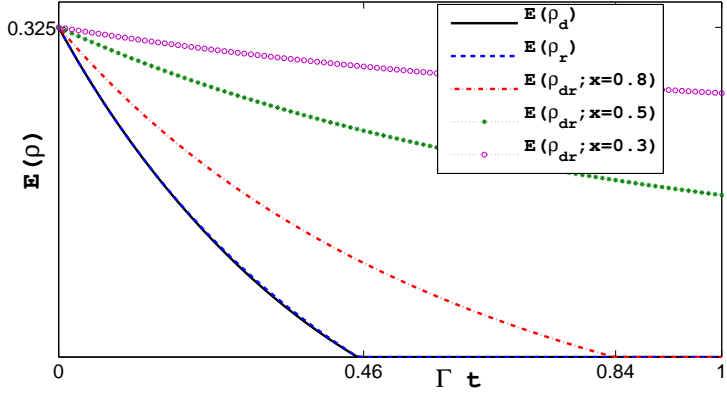}}
\centering
\caption{Genuine negativity for mixture of three-qubit $GHZ$ state plotted against parameter $\Gamma \, t$. The three curves are for 
double recovery operations.}
\label{Fig:R3QbGHZdr3}
\end{figure}

Finally, we consider the case where we apply recovery operations for two times. Once before the qubits interact with environment and 
once after that. This case we name as double recovery of genuine entanglement, denoted with a ``dr'' in subscript of density matrix 
in the figures and text. Figure~(\ref{Fig:R3QbGHZdr3}) show the effects of double recovery for mixture of $GHZ$ state. 
The solid (black) line is for amplitude damping without any recovery operations. The dashed (blue) line is for recovery operations applied for single 
time and after that qubits are exposed to amplitude damping. We see that these two lines almost overlap each other and both end at 
$\Gamma \, t \approx 0.46$. The other three curves are for decreasing values of parameter ``$x$'' and we see that as ``$x$'' decreases the recovery 
of genuine entanglement is considerably enhanced. Another interesting feature is that the initial amount of genuine entanglement remains same for all 
values of ``$x$''.  

\begin{figure}[t!]
\centering
\scalebox{2.25}{\includegraphics[width=1.99in]{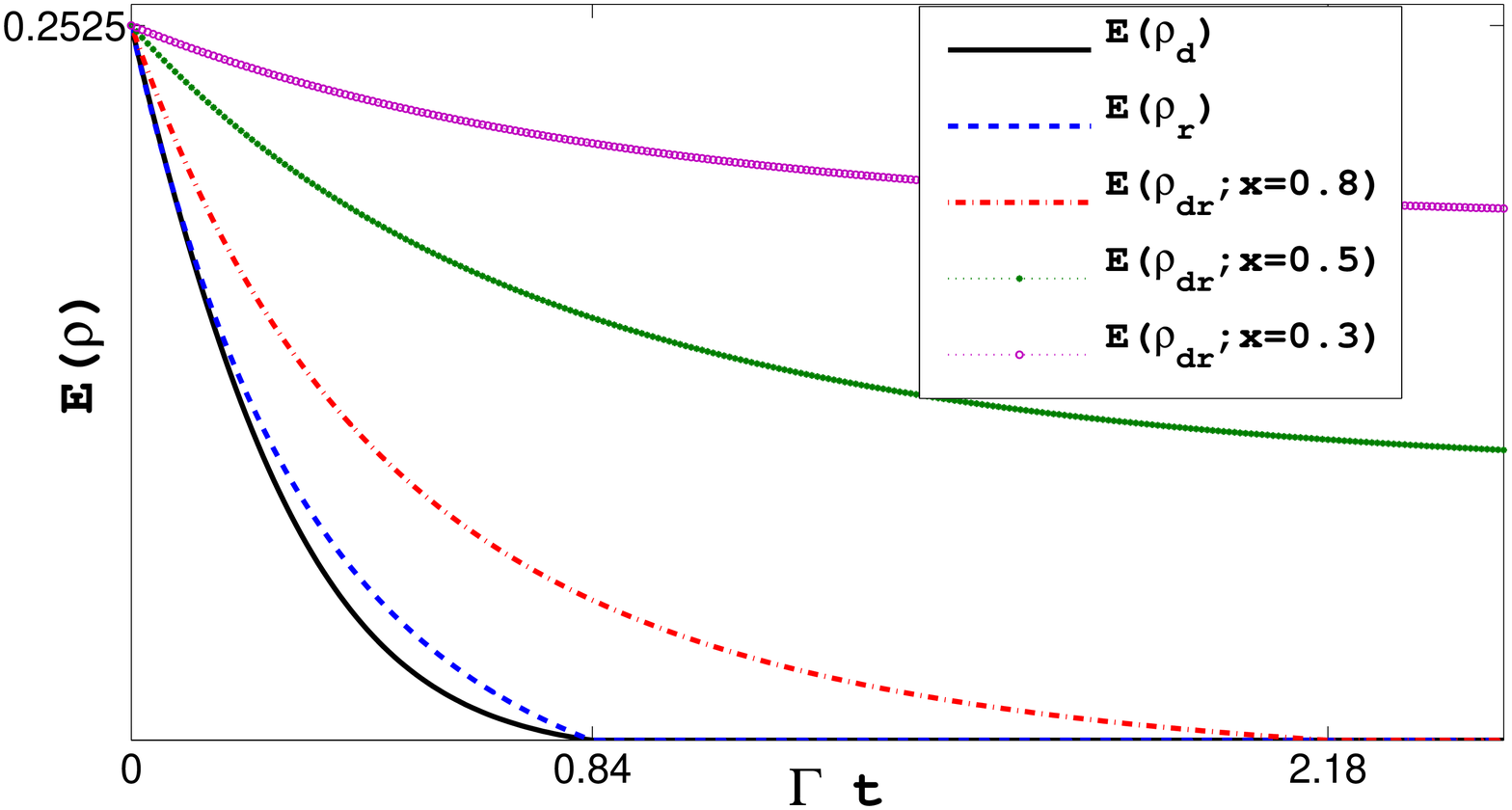}}
\caption{Genuine negativity for mixture of three-qubit $W$ state against parameter $\Gamma \, t$. The three curves for double recovery operations 
reflects the recovery of genuine entanglement.}
\label{Fig:R3QbWdr6}
\end{figure}
Figure~(\ref{Fig:R3QbWdr6}) shows the effects of double recovery operations for mixture of $W$ state. The solid line is for 
dynamics without any recovery oparations and dashed (blue) line is for operations applied before amplitude damping. Similar to mixture of 
$GHZ$ state, both curves reach to zero at $\Gamma t \approx 0.84$, however, here numerical value of genuine entanglement is enhanced and
can be seen more clearly. The other three curves show that recovery of genuine entanglement is greatly enhanced by decreasing the 
parameter ``$x$''.

\begin{figure}[t!]
\centering
\scalebox{2.25}{\includegraphics[width=1.99in]{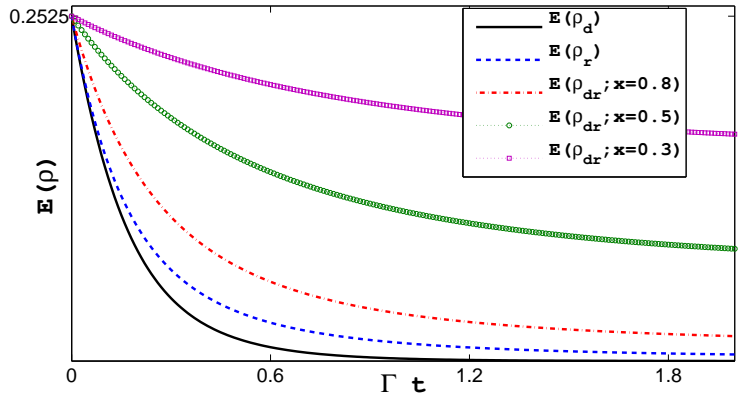}}
\caption{Genuine negativity for mixture of three-qubit $\ket{\widetilde{W}}$ state against parameter $\Gamma t$. 
The recovery of genuine entanglement is more clearly enhanced for this initial state.}
\label{Fig:R3QbWT7}
\end{figure}
Finally we check the behaviour for mixture of $\ket{\widetilde{W}}$ state (Eq.~(\ref{Eq:WT})).  
Figure~(\ref{Fig:R3QbWT7}) depicts the effects for amplitude damping without recovery operations (solid line), recovery operations only 
at the start (dashed blue line), and double recovery operations (other three curves). Interestingly, in this case, the single recovery 
operations enhance the degree of genuine entanglement more clearly than $W$ state. Also the other three curves shows that 
$|\widetilde{W}\rangle$ state is more robust against amplitude damping. This observation is interesting and surprising, as it has all 
three components as doubely excited as compared with $|W\rangle$ state. This state, we expected to be more fragile in the sense that it 
would loose its genuine entanglement more quickly, however, we have seen that its exhibiting quite opposite behaviour. The reason for this 
behaviour can be understood by examining the error matrix Eq.~(\ref{Eq:ERR}), in which we already discussed that for $\ket{\widetilde{W}}$ state, 
the error matrix entries are zero, meaning better recovery.

\section{Discussion and Summary}
\label{Conc}

We studied the effects of local recovery operations on genuine multiparticle entanglement for three qubit quantum states undergoing local 
amplitude damping. We analyzed several schemes in which such recovery operation could be applied. We found that for all recovery 
operations applied after the qubits are exposed to amplitude damping noise, there is no enhancement in the lifetime of genuine 
entanglement. We have studied the mixtures of W-type states and GHZ state with white noise as examples. We found that 
for all such schemes in which the recovery oparations are applied before, it is possible to prolong the 
lifetime of genuine entanglement upto some extent. For considerably and fully recovery of quantum states and genuine entanglement, 
the recovery operations must also need to be applied after the qubits are exposed to local amplitude damping. 
Only such double recovery scheme actually makes the quantum states robust against local amplitude damping. 
Our studies can be extended in several directions. First, it would be interesting to investigate the relation of time of applying 
second recovery operations with the decoherence time. In this work, we only consider the action of applying second operations at the start 
of the dynamics. It would be worth to figure out whether there is any critical time for applying such operations and after this time, it may 
not be possible to restore genuine entanglement. Second, it would be interesting to find out the similar recovery schemes for other types of 
decoherence models, like phase damping and/or depolarizing noise. 

\section*{Acknowledgements}

The author is grateful to referee for his/her constructive comments, which brought some clarity in the paper.

\end{document}